\documentclass[twocolumn,amsmath,amssymb]{revtex4}

\usepackage{graphicx}
\usepackage{dcolumn}

\usepackage{amsmath,amsthm}
\usepackage{amssymb}
\usepackage{epstopdf}
\usepackage{times}
\usepackage{mathrsfs}
\usepackage{color}
\usepackage{gensymb}
\usepackage{times}
\usepackage{subfigure}
\usepackage[]{graphicx}
\usepackage{amssymb}
\usepackage{amsmath}
\usepackage{bm}
\DeclareGraphicsExtensions{.png,.jpg,.pdf,.gif,.eps}
\usepackage{amsmath,amssymb,amsthm,amsfonts,eucal}
\usepackage[english]{babel}

\begin{document}
\title{A minimal integer automaton behind crystal plasticity}
\author{O\~guz Umut Salman$^{1}$}
\author{Lev Truskinovsky $^{1,2}$}
\affiliation{ $^{1}$LMS,  CNRS-UMR  7649,
Ecole Polytechnique, Route de Saclay, 91128 Palaiseau,  France\\
$^{2}$SEAS, Harvard University, 29 Oxford Street,
Cambridge, MA 02138, USA}

\date{\today}
\begin{abstract}
Power law fluctuations and scale free spatial patterns are known to characterize steady state plastic flow in crystalline materials. In this Letter we study the emergence of correlations in a simple Frenkel-Kontorova (FK) type model of 2D plasticity which is largely free of arbitrariness, amenable to analytical study and is capable of generating critical exponents matching experiments. Our main observation concerns the possibility to reduce continuum plasticity to an integer valued automaton revealing inherent discreteness of the plastic flow.
\end{abstract}
\maketitle

At the macroscale one usually assumes that crystalline materials flow plastically when averaged stresses exceed yield thresholds. At the microscale plasticity evolves through a sequence of slow-fast events involving collective pinning and depinning of dislocational structures. Classical engineering theory has been very successful in reproducing the most important plasticity phenomenology such as yield, hardening and shakedown \cite{Lubliner:1990cr}, however, a fully quantitative link between the phenomenological theory and the microscopic picture of plasticity remains elusive. The main reason is that the phenomenological approach implies spatial and temporal averaging in the system with poorly understood long range correlations.

The presence of such correlations have been confirmed by numerous experiments revealing  intermittent character of plastic activity with power law statistics  of avalanches and  self similar structure of dislocation cell structures \cite{Weiss:1997vn}. The emergence of power laws suggests that in plasticity the relation between the microscopic and the macroscopic models is more akin to turbulence than to elasticity \cite{Cottrell:2002zr, Papanikolaou:2010}.
Similar critical features of stationary nonequilibrium states have been observed in a variety other driven systems with threshold nonlinearity and rate independent dissipation including, for instance, tectonic faults,  magnets, and superconductors \cite{0034-4885-62-10-201}. The origin of scale free attractors in such systems is a subject of active research, in particular, the problem of classifying the universality classes remains largely open \cite{ISI:000086325400004}. In this situation finding the minimal representation of each class which is amenable to rigorous analysis is of significant general interest.

The experimental evidence of plastic criticality has been corroborated by several numerical models \cite{ISI:000237065700003}. The two main microscopic approaches are discrete dislocation dynamics, accounting for dislocation interactions on different slip planes  \cite{Miguel:2001dk, Miguel:2002dk,Devincre}, and a pinning-depinning model dealing with plasticity on a single slip plane  \cite{PhysRevB.69.214103, PhysRevLett.93.125502}. Different meso-scopic continuum
 models implying partial averaging have also been shown to generate power law statistics of avalanches with realistic exponents  \cite{PhysRevLett.78.4885, Papanikolaou:2010}. Since scale
 free dislocation activity is expected to be independent of either microscopic or macroscopic details, one can try to maximally simplify the underlying physics while still
 capturing the observed exponents and even characteristic shape functions \cite{Papanikolaou}. Presently the only analytically tractable models of plasticity are
 the mean field theory \cite{dharMajumdar1990,Zaisermoretti2005}, the FRG models of elastic depinning \cite{PhysRevE.79.051106} and
 the Abelian automata of sand pile type \cite{Dhar19994}. In this group only the automata models have a potential of capturing the
 whole complexity of the dislocation patterning and the goal of this Letter is to propose a formal reduction of a realistic plasticity model to a spin model with discrete time evolution. Instead of straightforward time discretization of continuum dynamics \cite{Narayan} we search for \emph{inherent} temporal discreteness   hidden behind the conventional gradient flow dynamics \cite{Trusk}.

 While the 1D automata, describing successfully plastic hysteresis and rate independent dissipation fall short of
 capturing plastic criticality \cite{Trusk},  the 2D
  automaton based models may be already adequate at least for FCC and hexagonal crystals; it is also noteworthy that intermittency has been mostly observed under single slip conditions \cite{Zaisermoretti2005, PhysRevLett.105.085503}.
In such cases one can get a realistic model by assuming that plasticity proceeds through the motion of a set of parallel edge dislocations.  We further neglect vectorial nature of the problem  and reduce the crystal to an array of coupled FK chains \cite{suziki1988}.
   In contrast to more conventional DDD modeling \cite{Miguel:2002dk} where nucleation and propagation \emph{rules} are not associated with the \emph{same} thresholds, the FK type models describes adequately both the multiplication of the defects and their kinetics including the finite size of the Peierls stress \cite{PhysRevLett.90.135502}.

\begin{figure}
\begin{center}
\includegraphics[scale=0.15]{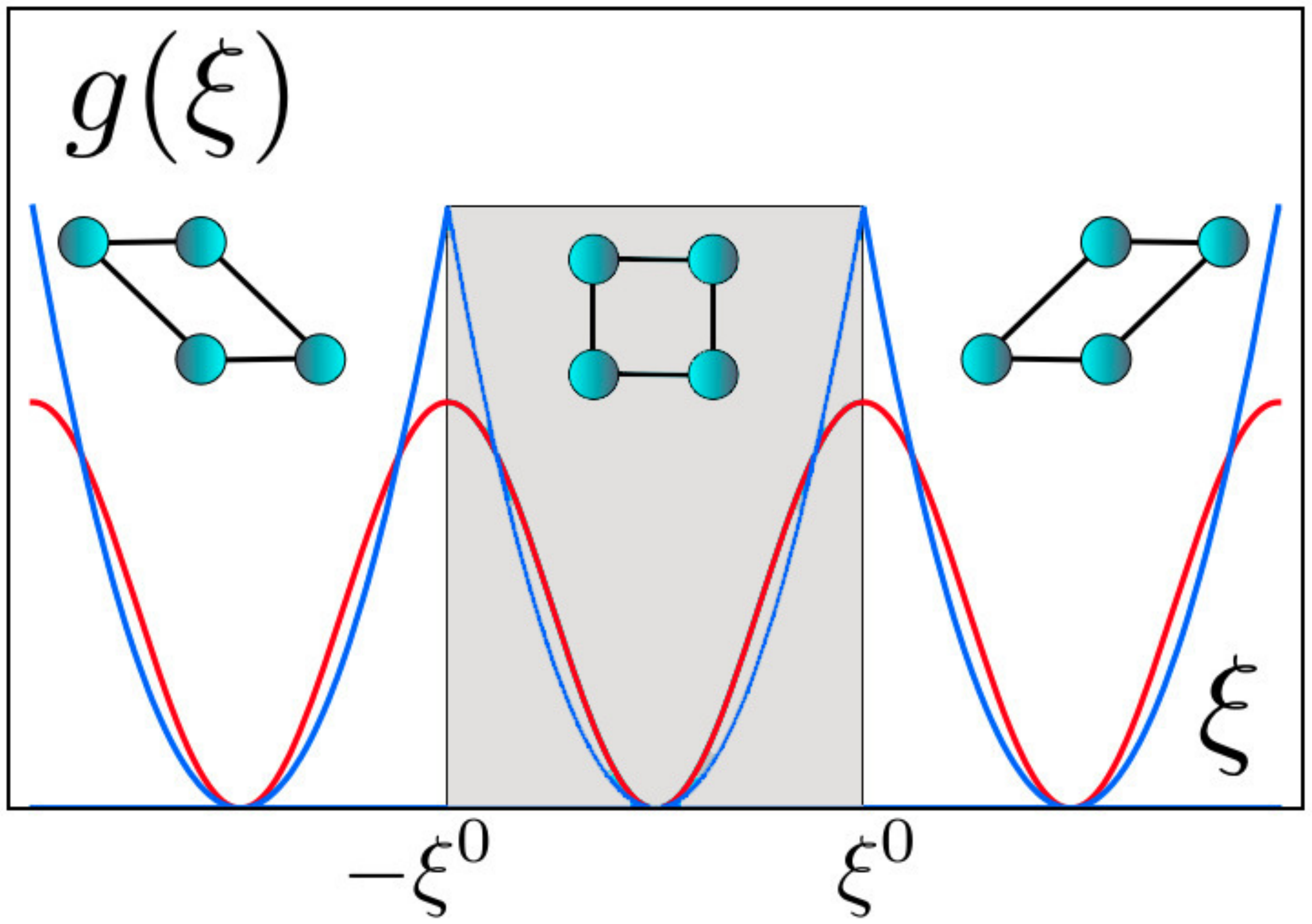}
\end{center}
\caption{\label{threshhold}\small Periodic dependence of energy on shear strain. Each minimum corresponds to a slip between adjacent atomic planes by one lattice spacing. Dislocations correspond to the boundaries of slipped regions. Red: continuous potential. Blue:  Piece-wise quadratic potential with the same linear elastic moduli. One elastic domain is shadowed.}
\end{figure}

To describe the model we first recall that classical continuum dislocation mechanics deals with the energy
$
\Phi(u)=\int \bar{\phi}(\nabla u)
$
where  $u(x)$ is the displacement field and the function $\bar{\phi}$ is quadratic. The displacement field is allowed to have finite discontinuities $[u]$ whose evolution is governed by phenomenological kinetic relations \cite{Kinetics}. The atomic structure of dislocations can be addressed by introducing  an internal length scale $a$ (Burgers parameter) and replacing continuum energy by  the discrete one which can be schematically represented as
$
\Phi(u)= a\sum \phi ([u]/a)
$
where now $u$ is a lattice field, $[u]$ is a discrete increment and the function $\phi$ is periodic with infinite symmetry group \cite{ISI:000222353800003}. Assuming that the order parameter is scalar, one can minimize out remaining linear strain variables: the simplest example of the resulting dressed description is the one dimensional FK model \cite{Braun}. Our minimal 2D setting can be viewed as an array of coupled FK chains with the energy \cite{suziki1988, PhysRevLett.90.135502}
 $\Phi(u)=\sum_{i,j} \phi(\theta,\xi),$
where $\theta(i,k)=u(i+1,k)-u(i,k)$  is an axial strain and  $\xi(i,k)=u(i,k+1) - u(i,k)$ is a shear strain; the displacement field $u(i,k)$ is defined on a $N \times N$ square lattice ($a=1$).  The potential $\phi$ is assumed to be quadratic in $\theta$ and periodic in $\xi$ (see Fig. \ref{threshhold}); to avoid synchronization we also added quenched disorder
 $
\phi(\theta, \xi)=  g (\xi )+ \frac{K}{2} (\theta)^2 -h_1 \xi-h_2 \theta.
 $
where $h_{1,2}(i,j),$ are independent gaussian random variables. Observe that the variables $\theta$ and $\xi$ are not independent and the long-range interactions can be revealed through minimizing out the non order parameter variable $\theta$  \cite{Saxena}.

The dynamic equations are taken in the form of the overdamped
gradient flow $\nu \dot{u}=- \partial \Phi(u)/\partial u,$ where $\nu$ is the ratio of the internal time scale and the time scale of the driving \cite{Trusk}. The driving in shear is performed through displacement controlled boundary condition $\sum _{k=0}^{N-1} \xi(i,k) =t$, where $t$ is the slow time playing the role of loading parameter; in longitudinal direction we assume periodic boundary condition $\sum_{i=0}^{N-1} \theta(i,k) =0$. In our numerical experiments we took $K=2$, $N=512$ and $g(\xi) = (2\pi)^{-2}(1-\cos(2\pi\xi))$. The initial state was dislocation free and the dispersion of disorder varied in the range $0.01-0.2$. For computations we used  an implicit-explicit FFT method \cite{ISI:A1994NU05400001}.
\begin{figure}[ht!]
\begin{center}
\subfigure[]{\label{m1}\includegraphics[height=3cm,width=4.0cm]{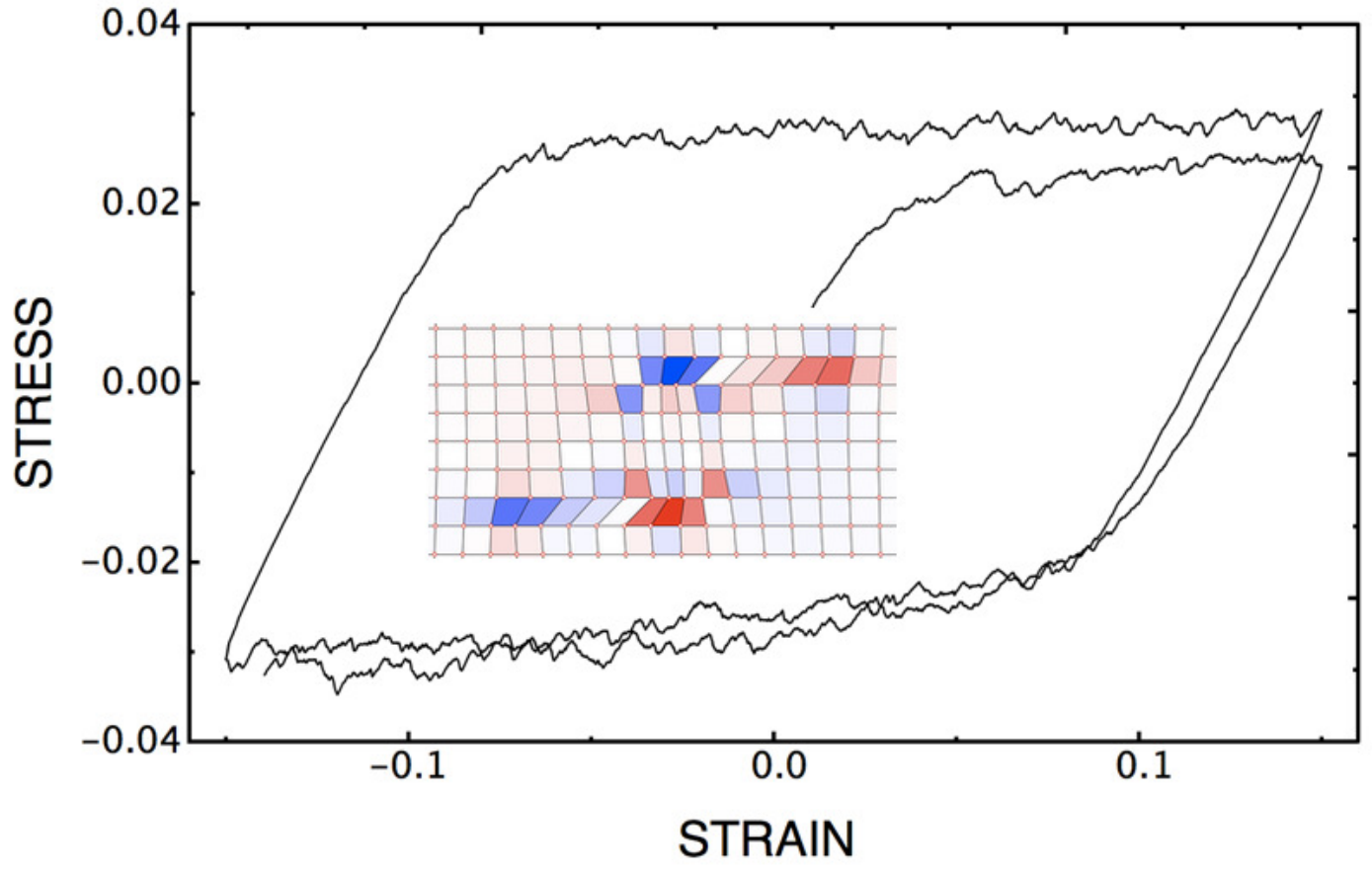}}
\subfigure[]{\label{m2}\includegraphics[height=3cm,width=4.3cm]{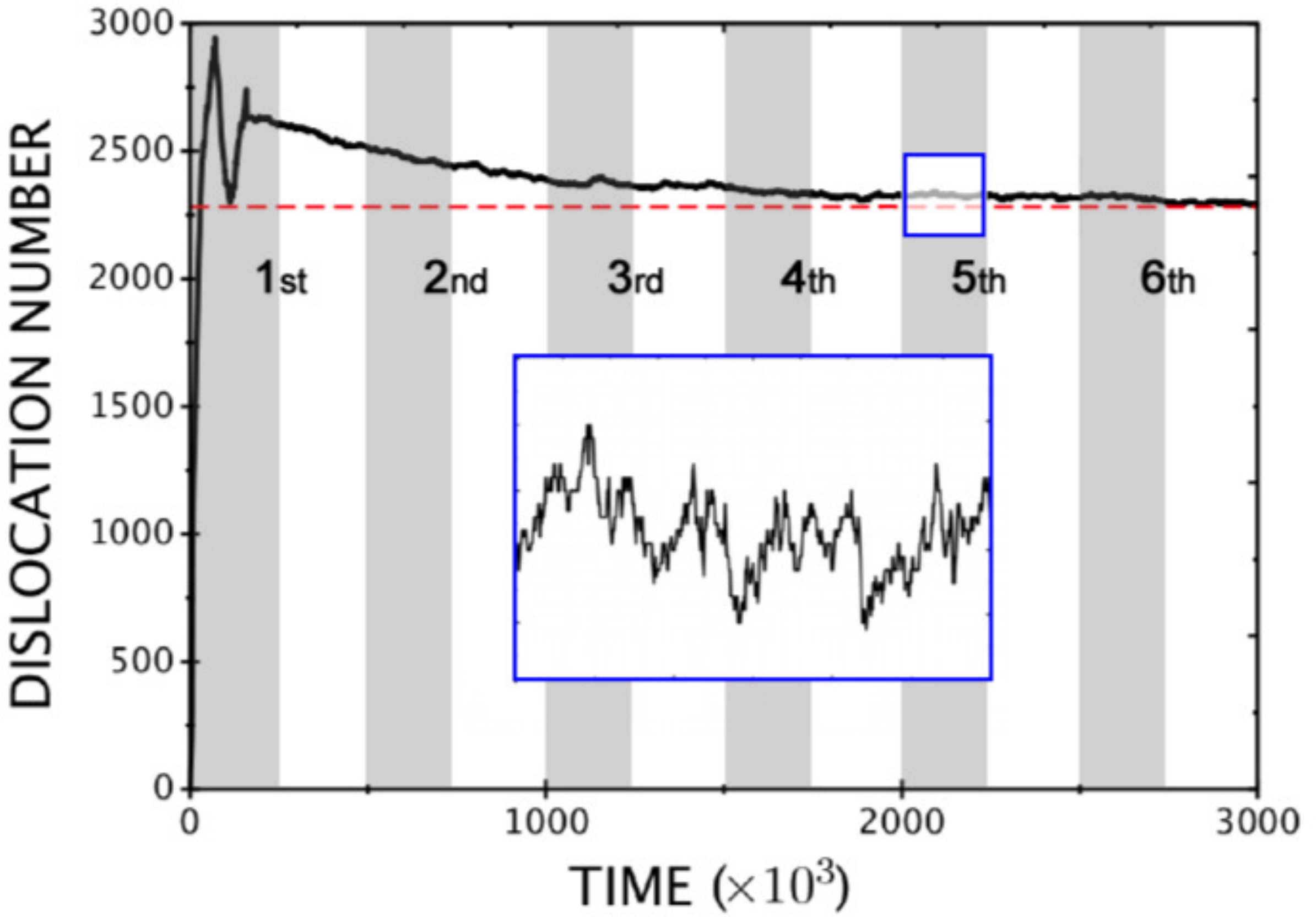}}
\end{center}
\caption{\label{atomic} (a) Macroscopic strain-stress curve showing plastic hysteresis and shakedown. An insert illustrates  a fragment of the deformed
lattice with two dislocation dipoles nucleated around an imperfection: red and blue colors correspond
   to dislocations of different sign. (b) Evolution of dislocation density with cycling. An insert shows small scale fluctuations.}
\end{figure}
\begin{figure}[ht!]
\begin{center}
\includegraphics[height=3.5cm,width=5.0cm]{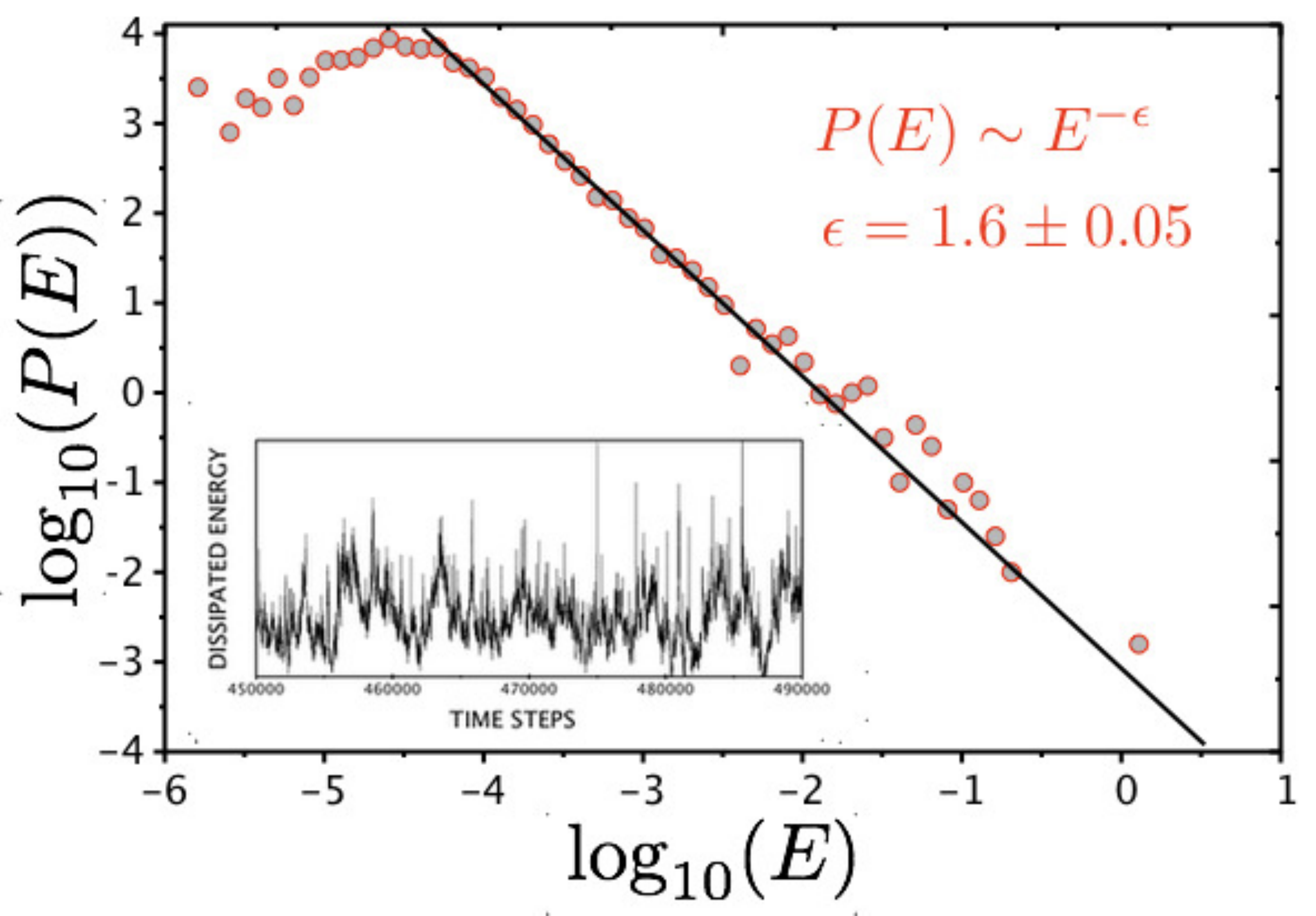}
\end{center}
\caption{\label{in_final_cycle} Probability distribution of dissipated energy in a steady state; an insert shows the structure of fluctuations during a typical cycle.
  }
\end{figure}
%
%
%
%

The results of direct numerical simulations (automaton reduction is discussed later) are presented in Fig. \ref{m1} where we show the macroscopic strain-stress curves covering first two cycles of loading/unloading in the hard device. Notice that hysteresis loops converge indicating that the system exhibits plastic shakedown. Reaching  steady state is marked by the stabilization of dislocation density, which also shows a characteristic nucleation related overshoot (see Fig. \ref{m2}). Steady state yielding is characterized by the formation of stable dislocation  structures (cells) with plastic activity limited to intermittent dislocational exchanges between the clusters; the latter   remain mostly stable from one cycle to another but have a finite lifetime as in observations \cite{jacobsen_science}. To separate individual avalanches we introduce an irrelevant threshold and define the avalanche energy by integrating viscous dissipation over its duration:
$E = N^{-2}\sum_{i,j}\int \dot u^2dt.
$
We observe that the probability distribution $P(E)$ stabilizes after several cycles (see Fig. \ref{in_final_cycle}) exhibiting a robust power law behavior with exponent $\epsilon\approx1.6\pm0.05$ obtained by maximum likelihood method \cite{clauset:661}. This value is in perfect agreement with experiments in ice crystals and fits the generally accepted range 1.4-1.6 \cite{Weiss:1997vn, Miguel:2001dk}; most remarkably it is also consistent with the value obtained for 2D colloidal crystals \cite{Pertsinidis}. The approximate proportionality
between the plastic slip size and the dissipated energy ensures that acoustic emission measurements would exhibit the same exponent $\epsilon$.

\begin{figure}
\begin{center}
\subfigure[]{\label{m3}\includegraphics[height=3.3cm,width=5.5cm]{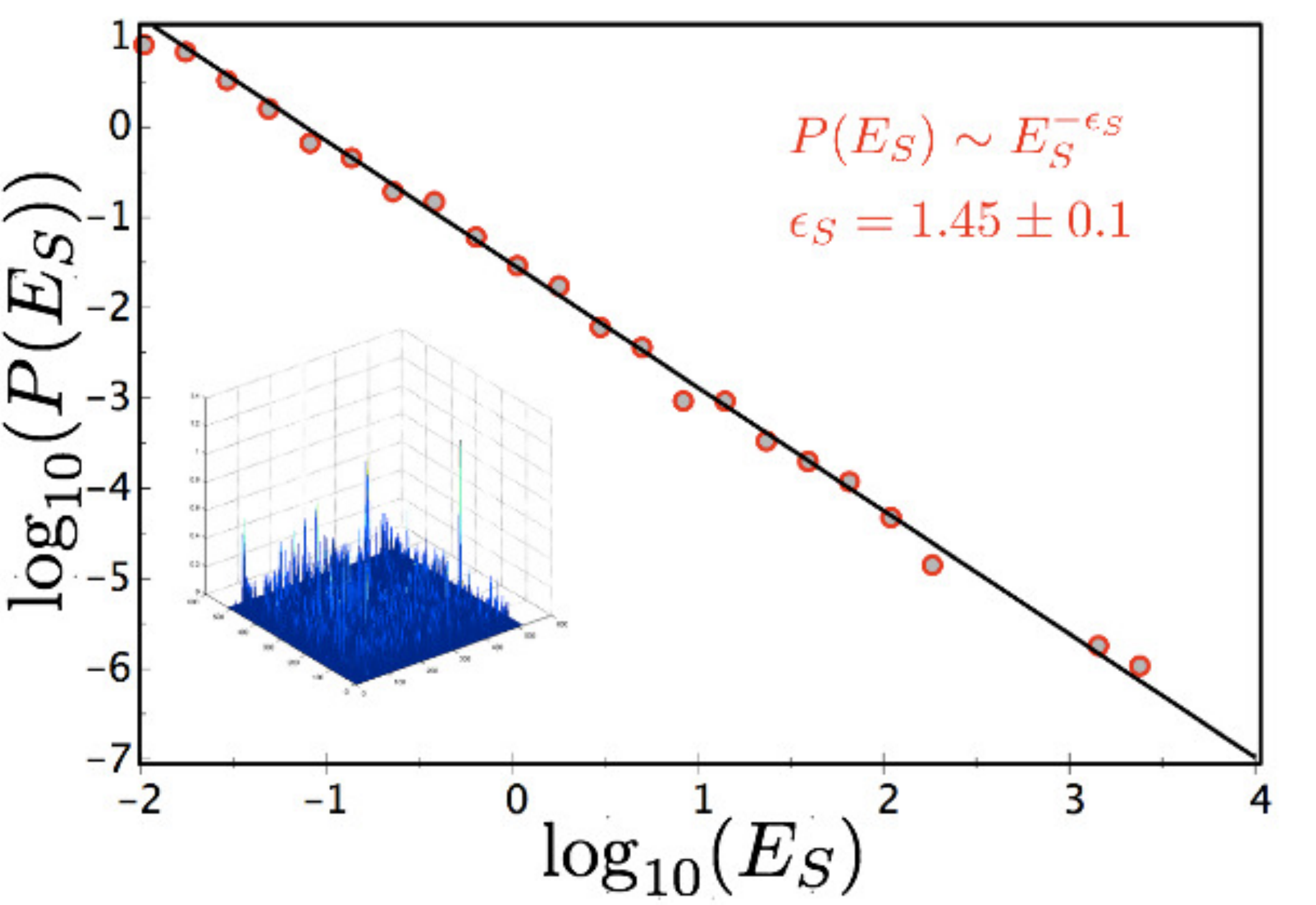}}
\subfigure[]{\label{corr_int}\includegraphics[height=3.3cm,width=5.6cm]{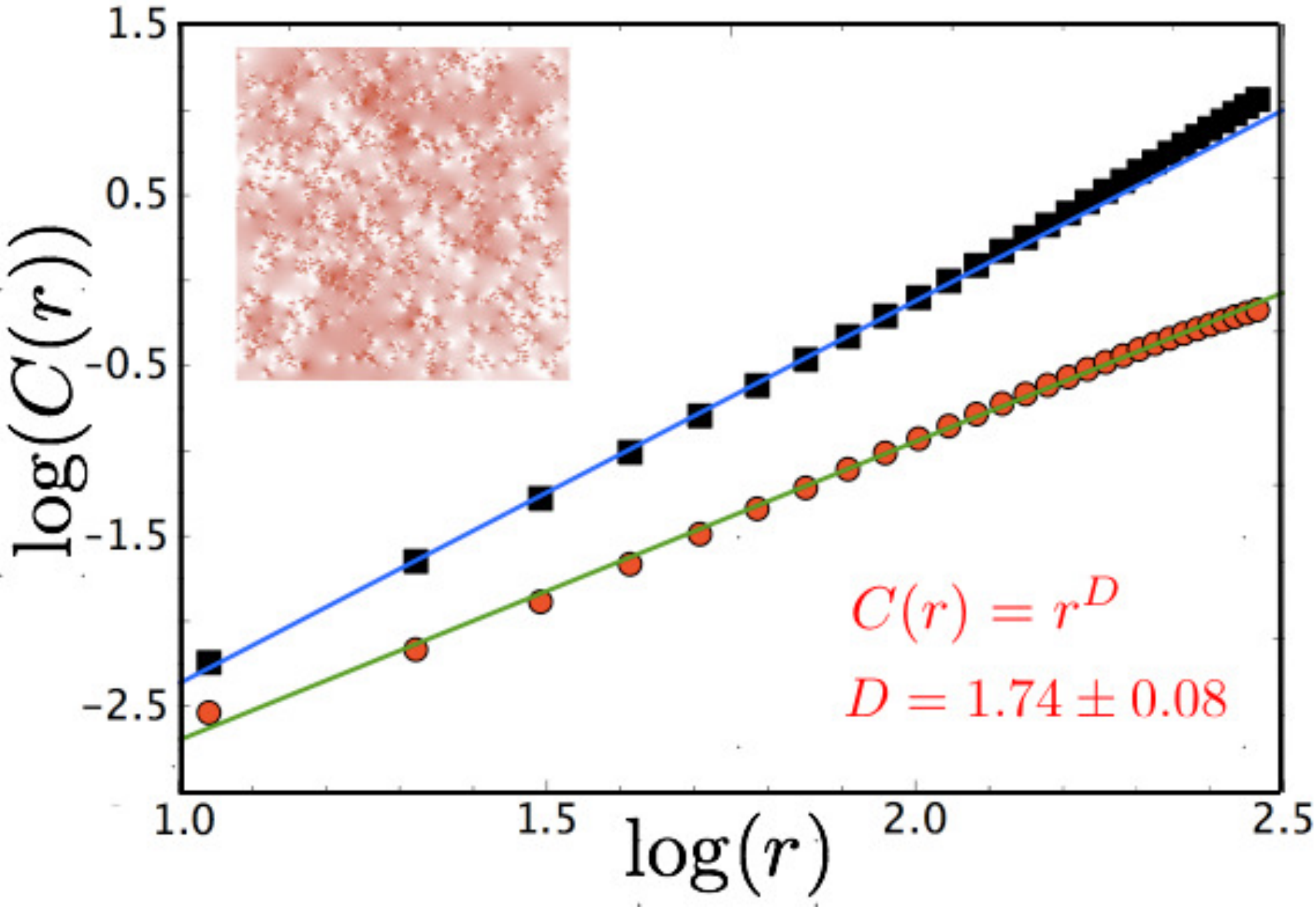}}
\end{center}
\caption{\small (a) Probability distribution of dislocation rich regions in the shakedown state; an insert shows spatial distribution of the energy density $\phi(\theta, \xi)$; (b) Correlation function $C(r)$ after the first cycle (squares) and after the fifth cycle (circles); an insert shows a characteristic stress field during steady yielding.}
\end{figure}
The spatial counterpart of the observed time correlations is the  fractal structure of dislocational patterns. The dislocation rich regions (clusters) can be identified by the localized peaks of the energy density landscape (see insert in Fig.\ref{m3}) and the corresponding probability density shows  a power law structure with exponent $1.45\pm0.1$ (Fig.\ref{m3}). Another way to quantify the fractal clustering is to compute the correlation function of the dislocation distribution $C(r)\sim r^D$  \cite{Hentschel:1982cz}. We observe  that during the first loading cycle $D\sim2.0$, which is expected given the random nature of the quenched disorder. With cycling the long range correlations develop (see Fig. \ref{corr_int} and in
 the shakedown regime we record $D\sim1.74$ independently of initial disorder. Note that dislocation patterns
 with $D \approx1.64-1.79$ have been observed experimentally in crystals with multiple slip systems;
 in simulations with a single slip system fractal patterning has been previously linked to the possibility of dislocation multiplication \cite{PhysRevLett.84.1487}
 which is operative in our model.

%

Despite the conceptual transparency of the above model, the mechanism of reaching the critical regime remains obscure. The model, however, can be simplified further if we notice that in quasi-static limit $\nu\rightarrow 0$ the relaxation is instantaneous and the system remains almost always in equilibrium
$
 \partial \Phi / \partial u=0.
$
In order to solve equilibrium equations analytically we can replace the smooth periodic potential by a piece-wise quadratic potential (see Fig. \ref{threshhold}) defined in each period $((d-1)\xi^0,(d+1)\xi^0)$ as
$
g(\xi ) = \frac{1}{2}(\xi-d)^2,
$
where $d$ is the new integer-valued spin variable describing a quantized slip; since at given lattice field $d(i,j)$
 the  equilibrium equations are linear the strain field can be found analytically (cf. \cite{Koslowski}). The Fourier image $\hat \xi(\bold q)$ of the shear strain  $\xi(i,j)$ reads
$$
\hat \xi(\bold q) = (s^+_y(\bold q)s^-_y(\bold q)\hat d(\bold q) + \hat H(\bold q))/\hat \lambda(\bold q),
$$
where $ \bold q=(q_x,q_y)=(2\pi k/N, 2\pi l /N)$ is the wave number. Here we defined
  $\hat H(\bold q)=s^-_x(\bold q) s^+_x(\bold q) \hat h_1(\bold q) + s^-_y(\bold q) s^+_y(\bold q) \hat h_2(\bold q)$
  and $\hat\lambda(\bold q)=2K (\cos(q_x)-1) + s^-_y(\bold q)s^+_y(\bold q) $ where $s^{\mp}_a (\bold q) = \pm (1-\cos (q_a ) \pm i\sin (q_a))$ with ${a}=x,y$. Notice also that we control  the average shear strain $\hat \xi_0(\bold q) = t \delta (\bold q)$
     and that the quenched disorder becomes the source of the residual strain $ \hat \xi_h(\bold q)=\hat H(\bold q)/\hat \lambda (\bold q) $.

It is now straightforward to reformulate the model as an integer valued automaton.
 Observe that the variable $\Delta \xi=\xi-(\xi_0+\xi_h)$ representing shear strain fluctuations must be confined between the thresholds
 $
-\xi^0- \xi(h,t)  <\Delta \xi(i,j) <\xi^0 - \xi(h,t) ,
 $
where $\xi(h,t)=[\hat \xi_h]_{\bold q}^{-1}+t$ and $[\cdot]_{\bold q}^{-1}$ denotes the inverse Fourier transform.
When $\Delta \xi$ reaches the threshold, the integer parameter  $d$ is updated $d\rightarrow d + M(\Delta \xi)$, where
$$
   M(\Delta \xi ) =\left\{
  \begin{aligned}
&{}  +1,\hspace{1mm}\text{if} \hspace{1mm} \Delta \xi >\xi^t -\xi(h,t),\\
&{} -1, \hspace{1mm}\text{if}\hspace{1mm} \Delta \xi  <-\xi^t -\xi(h,t) \\
&{}    \hspace{0.5cm} 0 \hspace{1mm}\text{otherwise.}
\end{aligned}
\right.
$$
After each increment of loading $t$ we recheck the stability until all the units are stabilized;  the dissipated energy during an avalanche is the difference of the total energies for two subsequent stable states.

\begin{figure}
\begin{center}
 \includegraphics[scale=0.25]{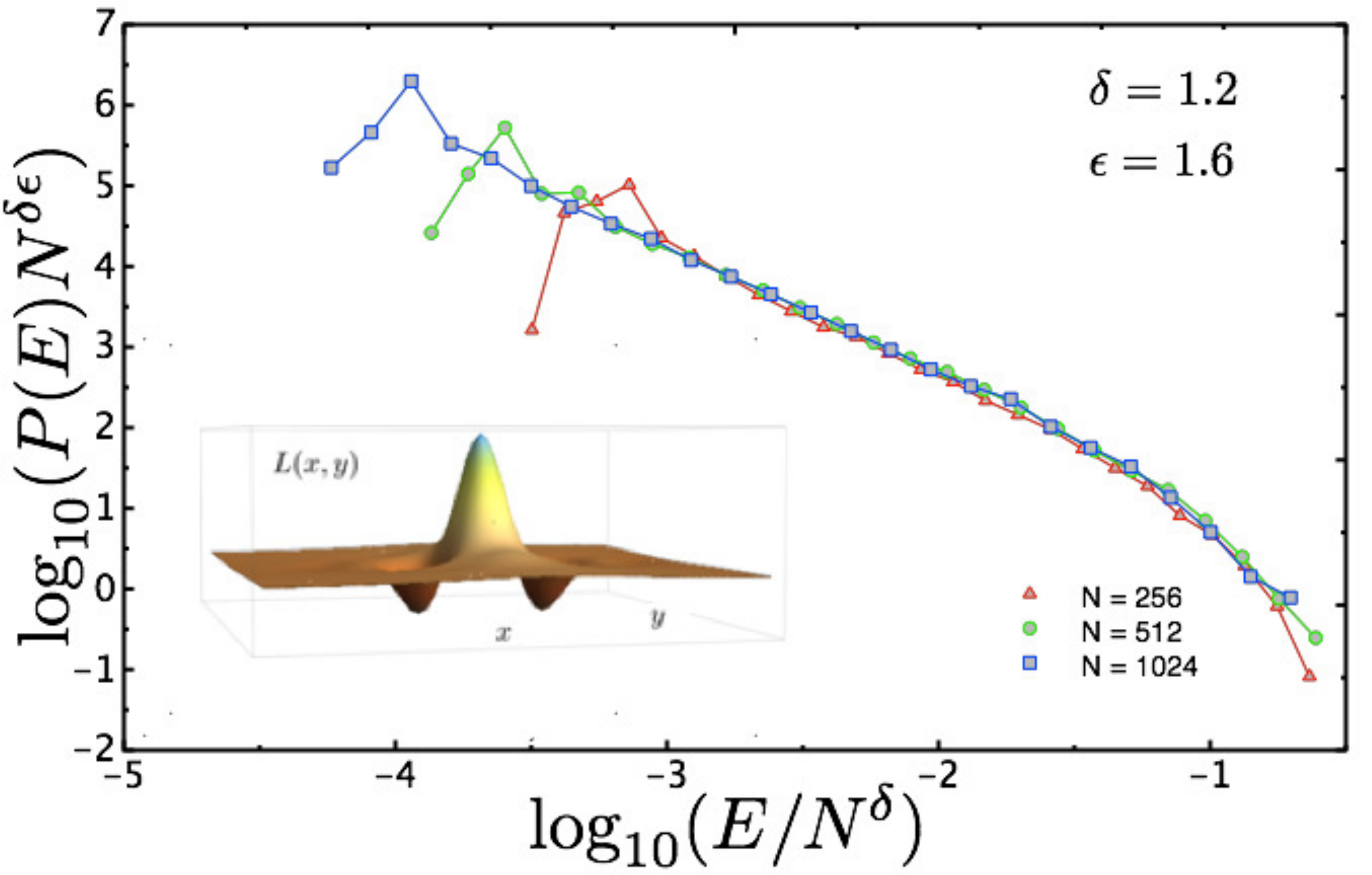}
\end{center}
\caption{\label{dc} \small Scaling collapse of the dissipated energy in automaton model for different system sizes $N$. An insert displays the kernel $L(x,y)$ and illustrates the update process in the automaton.}
\end{figure}

 The use of cellular automaton representation greatly reduces the complexity of numerical computations while
 the behavior of the system remains the same including the shape of the stress strain hysteresis, the evolution of the dislocation density and the structure of spatial and temporal correlations. To illustrate the statistics we show in Fig.  \ref{dc}
the finite-size scaling collapse of energy dissipation at the critical state; here we assumed that $P(E)=E^{-\epsilon}\varphi(E/E_c)$ with universal cut off function $\varphi$ and the cut off energy which diverges in the thermodynamic limit as $E_c\sim N^\delta$. Our computations show that again $\epsilon\approx1.6\pm0.05$ and that $\delta \approx 1.2\pm0.1$ which is close to the value $\delta \approx 1$ obtained for plastic strain increments in \cite{Miguel:2001dk, nikitas}.  These exponents are insensitive to the degree of disorder in the studied range; for larger disorder we observed a cut-off which is no longer size dependent.  Based on our computations we conclude that both models, the one with continuous dynamics and smooth potential, and the one with discrete dynamics and piece wise quadratic potential belong to the same universality class. This behavior is markedly different from the prediction of the mean field theory where smooth and cusped potentials lead to different universality classes  \cite{narayan2}.

An important question is whether the toppling rules in our automaton are Abelian meaning that the outcome of the instability in multiple sites does not depend on the toppling order. The update of the "slope" field $\Delta \xi$ can be represented in Fourier space as
$
\hat \Delta \xi  \rightarrow  \hat \Delta \xi - \hat L(\bold q) \hat M(\Delta \xi) ,
$
where
 $$
\hat L(\bold q) = \frac{\sin(q_y/2)^2}{\sin(q_y/2)^2+ K\sin(q_x/2)^2}
$$
is the analog of the toppling matrix in the sand-pile models.
The corresponding kernel $L(x,y)$ in the real space is highly anisotropic, long-range and conservative (see insert in Fig. \ref{dc}).
%
%
We compared numerically all conventional updating strategies and found that the microscopic configuration shows some small dependence on the choice of the strategy while the macroscopic observables, including the shakedown hysteresis loop and the statistics of avalanches (critical exponents) remain unaffected. One can conclude that our automaton has a weak (statistical) form of Abelian symmetry which may still be helpful for the mathematical analysis \cite{Dhar19994}.
Another important property of our automaton model is that it necessarily lowers the energy during each avalanche. The dissipative structure is obvious in the continuum  model and is inherited by the automaton model.

In conclusion, to elucidate the origin of self organized criticality in plasticity
 we reduced a realistic continuous dynamics to an integer automaton by replacing the fast
 dissipative stages with jump discontinuities controlled by random thresholds.
   The fact that despite the long range character of elastic interaction the
   computed exponents are different from the predictions of the mean field theory may mean that at least for some crystal classes plasticity is effectively a 2D phenomenon laying below the upper critical dimension.

The authors would like to thank A. Constantinescu,  S. Conti, A. Finel, S. Papanikolaou, F. J. Perez-Reche, S. Roux, E. Vives, J. Weiss and S. Zapperi and an anonymous reviewer for helpful discussions. 

%
%
%
%
\bibliographystyle{tPHM}

%
%
%
%
%
%
\end{document}